\documentclass[%
 aps,amsmath,amssymb,superscriptaddress,
 reprint]{revtex4-1}

\usepackage{graphicx}
\usepackage{dcolumn}
\usepackage{bm}
\usepackage{color}

\usepackage{titlesec} 
\titlespacing\section{0pt}{10pt}{4pt}
\titlespacing\subsection{0pt}{10pt}{2pt}

\begin{document}
\title{Absolute chiral sensing in dielectric metasurfaces with signal reversals}%
\author{Sotiris Droulias}\email{sdroulias@iesl.forth.gr}
\affiliation{Institute of Electronic Structure and Laser, FORTH, 71110 Heraklion, Crete, Greece}
\author{Lykourgos Bougas}\email{lybougas@uni-mainz.de}
\affiliation{Institut f\"ur Physik, Johannes Gutenberg Universit\"at-Mainz, 55128 Mainz, Germany}
\date{\today}%

\begin{abstract}
Sensing molecular chirality at the nanoscale has been a long-standing challenge due to the inherently weak nature of chiroptical signals, and nanophotonic approaches have proven fruitful in accessing these signals. However, in most cases, absolute chiral sensing of the total chiral refractive index has not been possible, while the strong inherent signals from the nanostructures themselves obscure the weak chiroptical signals. Here, we propose a dielectric metamaterial system that overcomes these limitations and allows for absolute measurements of the total chirality, and the possibility for a crucial signal reversal that enables chirality measurements without the need for sample removal. As proof of principle, we demonstrate signal-enhancements by a factor of 200 for ultrathin, sub-wavelength, chiral samples over a uniform and accessible area.
\end{abstract}
\maketitle

\indent Chirality, a geometric property in which an object is non-superimposable with its mirror image, is a fundamental property of life with far-reaching implications among many research disciplines. Most notably, a molecule's chirality dictates its function\,\cite{Eliel2008}, particularly its metabolic uptake and pharmacological effects, such as its potency and toxicity\,\cite{Francotte2006,Hutt1996,Nguyen2006}. Sensing chirality is, therefore, of fundamental importance for research in analytical chemistry, biology, and pharmacology, and is routinely applied in the agricultural, pharmaceutical and chemical industries for enantiopurity/quality control\,\cite{Busch2006}.\\
\indent Nanophotonic approaches have proven to be a powerful means for granting access to weak chiroptical signals not previously attainable with traditional polarimetric techniques\,\cite{Hendry2010,Maoz2013,Zhao2017,Tullius2015,Tullius2017,Garcia2018,Kneer2018}. However, in most implementations, the employed nanosystems have intrinsic chiroptical responses that contribute to the total signal, often precluding the absolute measurement of chirality (handedness and magnitude). Importantly, these approaches do not allow for quantitative detection of both the real and imaginary part of the refractive index of a chiral substance (responsible for refraction and absorption, respectively)\,\cite{Hendry2010,Maoz2013,Tullius2015,Zhao2017,Garcia2018,Kneer2018,Mohammadi2018,Solomon2019}, a crucially sought after aspect for any chiral sensing technique. Nanophotonic systems are unique platforms for the development of lab-on-a-chip chiral-sensing devices\,\cite{Nagl2011}, but it is vital to design them such that they overcome the above mentioned limitations and yield accurate results.\\
\begin{figure*}[ht!]
		\includegraphics[width=\linewidth]{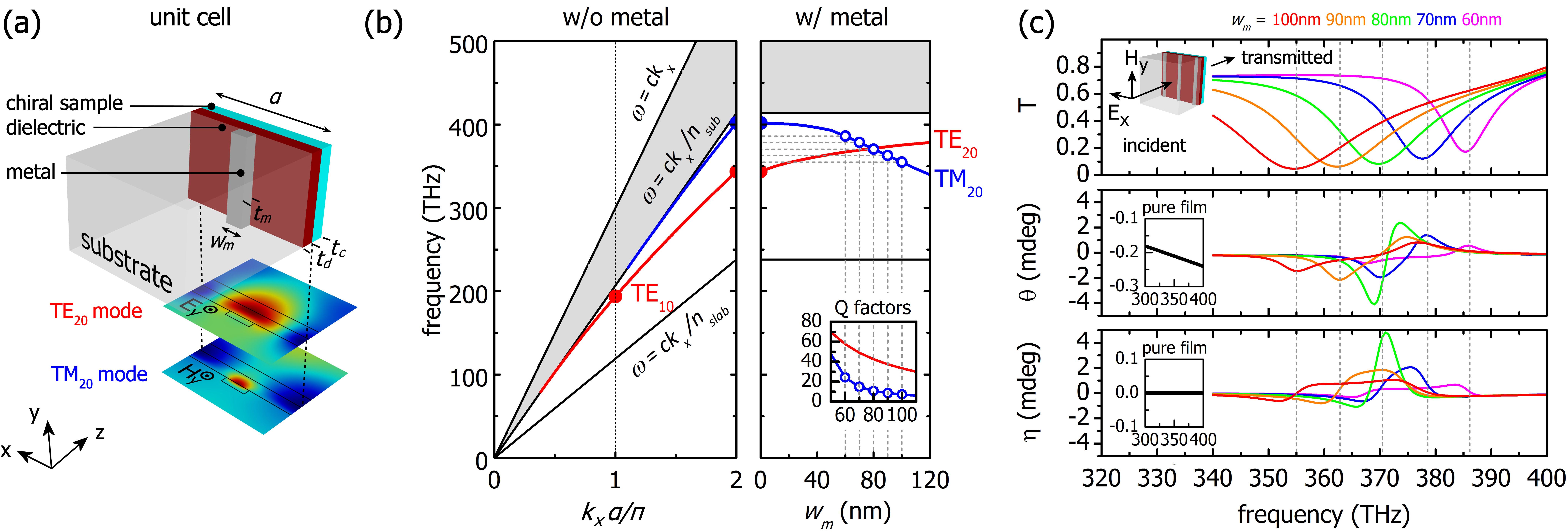}\vspace*{-10pt}
	\caption{\small{Dielectric metasurface for enhanced chiral sensing. (a) Schematic of the metasurface illustrating one unit cell (periodicity $\alpha$) (b) \textit{Left panel}: dispersion of the lowest order TE$_0$ (red line) and TM$_0$ (blue line) modes of the dielectric slab in the absence of metal ($\alpha = 500$\,nm, $n_{\rm{slab}} = 2.52$, and $n_{\rm{sub}} = 1.45$). \textit{Right panel}: spectral tuning of the TE$_{20}$ and TM$_{20}$ modes, as a function of the metal width, $w_m$ (inset: Q-factors). (c) Transmittance spectra and enhanced far-field chiroptical signals (optical rotation $\theta$ and ellipticity $\eta$) as a function of TE$_{20}$\,-\,TM$_{20}$ spectral alignment under TM excitation for a system with $\kappa=+10^{-5}$. The TM-polarized incident wave excites the TM$_{20}$ mode and the chiral layer provides the coupling with the TE$_{20}$ mode.}}
    	\label{fig:fig1}
\end{figure*}
\indent In this Letter, we present a metamaterial system that offers critical advantages over alternative nanophotonic-based chiral sensing techniques: (a) absolute measurement of chirality and quantitative detection of both the real and imaginary part of the refractive index of a chiral substance; (b) enhanced chiroptical responses for versatile excitation modalities; and (c) the introduction of a crucial signal reversal that suppresses undesired achiral signals originating from the metamaterial system without requiring sample removal for a null-sample measurement, while also enabling the use of modulation techniques for increased detection sensitivities. \\
\indent The principle of operation is based on appropriately tailoring two modes of the metasurface that provide strong dipole moments, one of electric type and one of magnetic type. As chiral matter-wave interactions require a nonvanishing pseudoscalar product between $\mathbf{E}$ and $\mathbf{B}$ ($\mathbf{E}$ and $\mathbf{B}$ are the time-dependent electric and magnetic fields, respectively), the combined action of the two modes enables the strong, near-field coupling between the metasurface and the chiral substance. As we demonstrate, excitation of either mode is capable of providing strong chiroptical far-field signals that can be further enhanced by the metasurface's anisotropy, without relying on the excitation of superchiral local fields\,\cite{Tang2010}; that is, superchirality, although beneficial, is not a prerequisite for our system, in contrast to most nanophotonic-based chiral sensing approaches\,\cite{Schaferling2012OSA,Schaferling2012PRX,Mohammadi2018,Poulikakos2018}.\\	
\indent A schematic of our metamaterial system is shown in Fig.\,\ref{fig:fig1}\,(a). The system consists of a thin dielectric slab that supports TE (components $H_x$, $E_y$, $H_z$) and TM (components $E_x$, $H_y$, $E_z$) waveguide modes with continuous dispersion [Fig.\,\ref{fig:fig1}\,(b)]. To implement the electric/magnetic-moment pair we work with TE$_0$ and TM$_0$, the lowest order waveguide modes, and utilize their dominant field components, i.e. $E_y$ and $H_y$, respectively. In order to spatially quantize them and achieve discrete sets of resonant states we place metallic wires on the slab with periodicity $\alpha$ (oriented parallel to the $y$-axis) that result in: (a) pinning the field distribution within each unit cell of length $\alpha$, (b) turning the system from continuous to periodic, thus, giving rise to a band structure with periodicity $\pi/\alpha$, and (c) splitting each of the two modes at the edge of the Brillouin zone into two discrete spatially shifted versions, so that their dominant-field component has either a node or an antinode at the metal region\,\cite{Droulias2017PRL,Droulias2017PRB,Droulias2018}. Because $E_y$ and $H_y$ belong to different modes, to maximize their interaction with the chiral system, the two modes must be aligned both (a) spectrally and (b) spatially. Spectral alignment is achieved by tuning the mode frequencies via the metal width, $w_m$ [Fig.\,\ref{fig:fig1}\,(b)], and spatial overlap by selecting to work with the TE$_0$ and TM$_0$ modes with antinodes at the metal region\,\cite{Christ2003,Floess2015,Droulias2017PRB,Droulias2018}. We refer to these modes as TE$_{20}$ and TM$_{20}$, respectively, as we operate at $k_x = 2\pi/\alpha$ (the in-plane wavenumber) [Fig.\,\ref{fig:fig1}\,(a)\&(b)]. As shown in Fig.\,\ref{fig:fig1}, their maxima occur at the metal and at the edges of the unit cell, and their strong near-fields extend outside the dielectric slab region enabling sufficient coupling with a chiral layer when placed on the slab; this is a significant advantage that renders the metasurface a large uniform sensing surface, contrary to schemes based on plasmonic hot spots\,\cite{Tang2013APL,Zhao2017,Tullius2017}. \\
\indent We focus on design parameters for chiral sensing in the visible and near-infrared (here we choose 800\,nm). The slab has refractive index $n_{\rm{slab}} = 2.52$ (e.g. TiO$_2$) and the metal is a Drude silver (Ag) of permittivity based on Johnson and Christy data\,\cite{JohnsonChristy}. The dielectric slab has thickness $t_d =80$\,nm and the metallic wires have periodicity $\alpha = 500$\,nm and thickness $t_m = 50$\,nm; their width $w_m$ is used for the spectral tuning of the TE$_{20}$, TM$_{20}$ eigenfrequencies [Fig.\,\ref{fig:fig1}\,(b)]\,\cite{Floess2015}. The chiral layer, which we place on the slab, has thickness $t_c = 50$\,nm, refractive index $n_c = 1.33-10^{-4}i$\,\footnote{We add artificial losses in the average refractive index of the chiral layer to ensure passivity when we examine cases of Im($\kappa$)$\neq0$.}, and chirality (Pasteur) parameter $\kappa = 10^{-5}$, which is a realistic value for the chirality parameter of chiral molecules (for, e.g., aqueous solutions of monosaccharides\,\cite{Sofikitis2014,Bougas2015} or biomolecules\,\cite{Abdulrahman2012}). The entire space above the chiral layer is water with $n=1.33$, while we place the whole system on a glass substrate of $n_{\rm{sub}} = 1.45$, which serves as a mechanical support. Our system is examined with full-wave vectorial Finite Element Method (FEM) simulations, utilizing the commercial software COMSOL Multiphysics.\\
\indent As illustrated in Fig.\,1\,(c), we excite the system at normal incidence from the substrate side with a linearly polarized wave. When the incident $H$-field is parallel to the wires, i.e. $\mathbf{H}=H_y\hat{y}$, it excites solely the TM$_{20}$ mode (components $E_x$, $H_y$, $E_z$), which cannot couple to the orthogonal TE$_{20}$  mode unless the chiral layer provides the necessary mode-coupling via $\kappa$ [similarly an incident field with $\mathbf{E}=E_y\hat{y}$ couples only to TE$_{20}$ (components $H_x$, $E_y$, $H_z$)]. Therefore, for $\kappa\neq0$ both modes are excited and any effect observed in the far-field is due to the chiral layer entirely, i.e. the signal is free from backgrounds from the photonic structure. To demonstrate this, we examine our system under \textbf{H}$\|\hat{y}$, which we refer as TM-illumination to emphasize the fact that this particular polarization directly couples with the TM$_{20}$ mode [see Supplemental Information (SI) for excitation with \textbf{E}$\|\hat{y}$ (TE illumination)]. In Fig.\,1\,(c) we plot the transmittance and analyze the polarization of the transmitted wave in terms of its rotation $\theta$ and ellipticity $\eta$, to detect the effect of chirality on the incident wave. Starting with $w_m$ = 60\,nm, as $w_m$ increases the TE$_{20}$-TM$_{20}$ spectral separation reduces resulting in increased $\theta$ and $\eta$ values in the far-field, which are maximized for $w_m = 80$\,nm where the two modes are spectrally aligned. Further increase in $w_m$ leads to detuning of the modes and hence weaker signals. This process is mediated entirely by the mode alignment since, as $w_m$ increases, the Q factors drop monotonically [Fig.\,\ref{fig:fig1}\,(b), right panel, inset], while $\theta$, $\eta$ evolve non-monotonically. For a 50\,nm thin chiral layer we obtain chiroptical rotation ($\theta$) signals as large as 6.5\,mdeg peak-to-peak, for a transmittance of $\sim$10\%. As a comparison, the optical rotation signal from a transmission measurement of a 50\,nm chiral layer with $\kappa=10^{-5}$ at 800\,nm is $\sim$0.24\,mdeg, achieving, thus, enhancements by a factor of $\sim$27 [Fig.\,\ref{fig:fig1}\,(c), insets].\\
\begin{figure}[t!]
		\includegraphics[width=\linewidth]{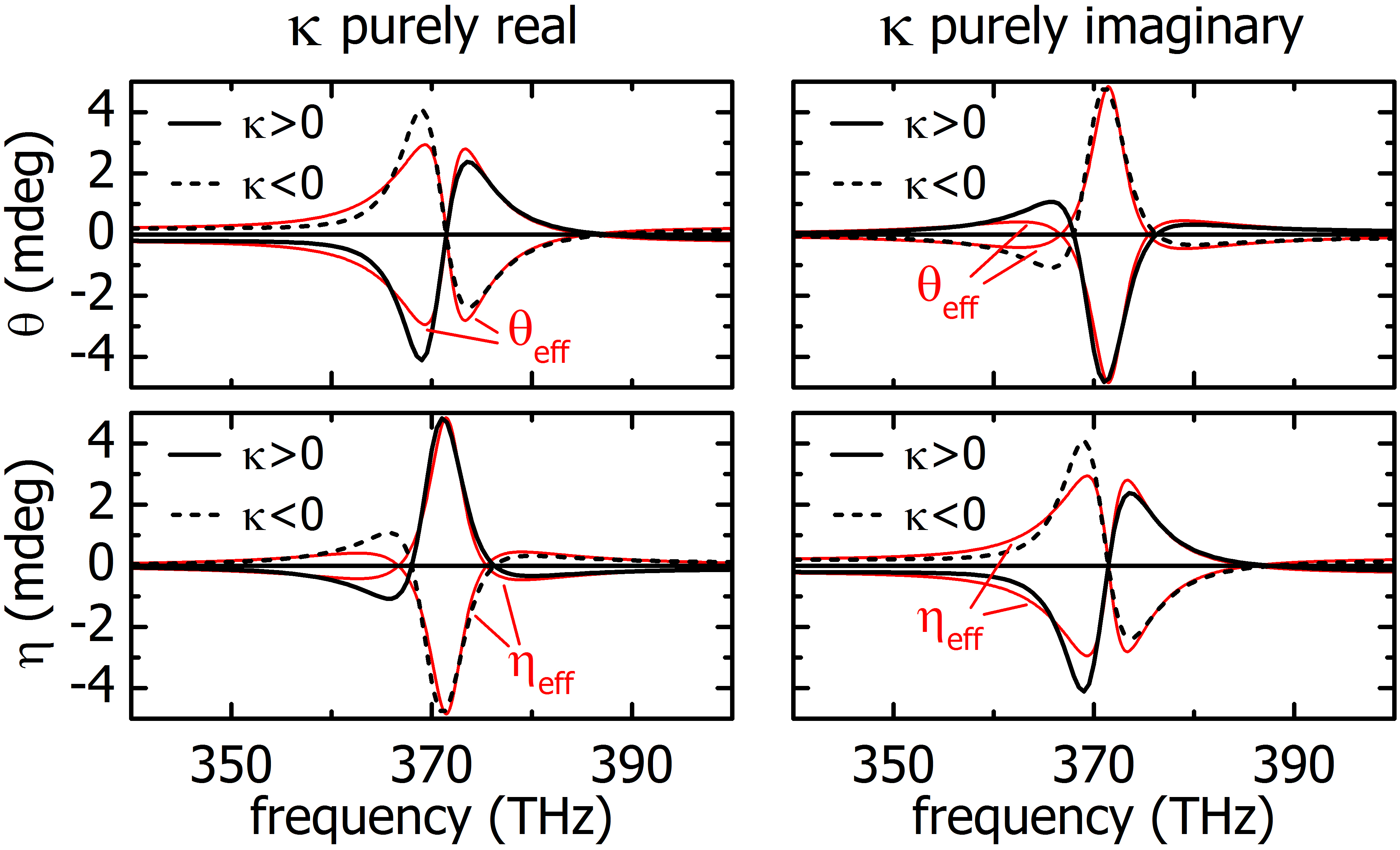}\vspace*{-8pt}
	\caption{\small{Enhanced absolute sensing of total chirality. Simulations of far-field signals of optical rotation $\theta$ and ellipticity $\eta$ for spectrally aligned modes under TM-illumination ($w_m=80$\,nm; Fig.\,1), and fit with analytical model ($\theta_{\rm{eff}}$ and $\eta_{\rm{eff}}$, respectively). \textit{Left column}: $\kappa=\pm10^{-5}$ (purely real). \textit{Right column}: $\kappa=\pm10^{-5}i$ (purely imaginary).}}
    	\label{fig:absolute}
\end{figure}
\indent To understand the observed enhanced signals, we model our thin, subwavelength system as an infinite polarizable sheet, which supports an electric- and a magnetic dipole moment, \textbf{p}$=p\hat{y}$ and \textbf{m}$=m\hat{y}$, respectively. In the actual system the incident field excites either TE$_{20}$ or TM$_{20}$, and the two modes are subsequently coupled via the chirality parameter $\kappa$. Accordingly, in the sheet model, $p$ and $m$ are driven by the external forces $f_{\rm{TE}}$ and $f_{\rm{TM}}$, respectively, and are coupled by a coupling constant $\kappa_c$. Hence, we can write a coupled oscillator model for $p$ and $m$ as (see SI for details)\,\cite{Tassin2012,Tullius2017}:\\
\begin{equation}\label{sheetmodel}
\begin{gathered}
\frac{d^2 p(t)}{dt^2}+\gamma_{\rm{TE}}\frac{d p(t)}{dt}+\omega_{\rm{TE}}^2 p(t)+\kappa_c m(t)=f_{\rm{TE}}(t)\\
\frac{d^2 m(t)}{dt^2}+\gamma_{\rm{TM}}\frac{d m(t)}{dt}+\omega_{\rm{TM}}^2 m(t)-\kappa_c p(t)=f_{\rm{TM}}(t),
\end{gathered}
\end{equation}
where $\omega_{\rm{TE}}$, $\omega_{\rm{TM}}$ are the resonant angular frequencies of the TE$_{20}$ and TM$_{20}$ modes and $\gamma_{\rm{TE}}$, $\gamma_{\rm{TM}}$ are their respective damping rates. In the frequency domain, assuming a solution of the form $p(t)=\tilde{p}(\omega)exp(-i\omega t)$ and $m(t)=\tilde{m}(\omega)exp(-i\omega t)$, the system takes the simple form:
\begin{widetext}
\begin{equation}\label{solutions}
\begin{gathered}
\tilde{p}(\omega)=\frac{ D_{\rm{TM}}(\omega) }{D_{\rm{TE}}(\omega)D_{\rm{TM}}(\omega) +\kappa_c^2 }\tilde{f}_{\rm{TE}}(\omega)+i \frac{ i\kappa_c}{D_{\rm{TE}}(\omega)D_{\rm{TM}}(\omega) +\kappa_c^2 }\tilde{f}_{\rm{TM}}(\omega) \\
\tilde{m}(\omega)=\frac{ D_{\rm{TE}}(\omega) }{D_{\rm{TE}}(\omega)D_{\rm{TM}}(\omega) +\kappa_c^2 }\tilde{f}_{\rm{TM}}(\omega)-i \frac{ i\kappa_c}{D_{\rm{TE}}(\omega)D_{\rm{TM}}(\omega) +\kappa_c^2 }\tilde{f}_{\rm{TE}}(\omega),
\end{gathered}
\end{equation}
\end{widetext}
where $D_n(\omega) = \omega_n^2 -\omega^2 -i\gamma_n\omega$, $n=\{\rm{TE}, \rm{TM}\}$. This result qualitatively describes the excitation of $p$ and $m$ under TE and/or TM incident polarization. For pure TM-illumination, as is the case in our simulations, $f_{\rm{TE}} = 0$ clearly leads to the excitation of both $p$ and $m$, despite the absence of a direct external excitation of $p$. Comparing the form of Eq.\,\ref{solutions} to the constitutive relations, i.e. $\rm{\textbf{D}} = \epsilon \rm{\textbf{E}} + i(\kappa/c)\rm{\textbf{H}}$, $\rm{\textbf{B}} = \mu \rm{\textbf{H}} - i(\kappa/c)\rm{\textbf{E}}$, the near-field contribution to the effective chirality of the composite system is given by:
\begin{equation}\label{kappaeff}
\kappa_s=a_c \frac{i\kappa_c}{D_{\rm{TE}}(\omega)D_{\rm{TM}}(\omega) +\kappa_c^2 }
\end{equation}
where $a_c$ is a constant that associates the surface susceptibility to the bulk effective chirality. As Eq.\,\ref{kappaeff} dictates, spectrally aligned modes of high Q factors (i.e. low $\gamma_{TE}$, $\gamma_{TM}$) lead to enhanced chiroptical signals. \\
\indent Besides providing physical insight, our simple analytical model can also qualitatively describe the overall response of our system. In Fig.\,\,\ref{fig:absolute} we use Eq.\,\ref{kappaeff} to fit the simulated results for the case of aligned modes, where the enhancement is maximum ($w_m$ = 80\,nm; Fig.\,\ref{fig:fig1}). We now set $\kappa$ = $\pm10^{-5}$ and $\kappa$ = $\pm10^{-5}i$ to investigate the cases of both chiral-dependent refraction and absorption. First, we demonstrate that the individual simulations show a clear distinction in the $\theta$ and $\eta$ signatures between the four cases, that is, our system provides absolute measurements of total chirality (both real and imaginary part of $\kappa$). Second, we verify this using our coupled oscillator model, which, despite its simplicity, successfully reproduces the numerical simulations. To fit the numerical simulations we use $\omega_{TE}=\omega_{TM}= 2\pi\times371\,\rm{THz}$, $\gamma_{TE}= 2\pi\times4.4\,\rm{THz}$, $\gamma_{TM}= 2\pi\times17.3\,\rm{THz}$ from the eigenmode simulations, and $a_c\kappa_c= \kappa\times(2\pi\times118\,\rm{THz})^4$, where $\kappa$ = $\pm10^{-5},\pm10^{-5}i$, with the approximation $D_{\rm{TE}}(\omega)D_{\rm{TM}}(\omega)+\kappa_{c}^2 \approx D_{\rm{TE}}(\omega)D_{\rm{TM}}(\omega)$ (see SI for details). The far-field chiroptical signals result from the contribution of both $\kappa_s$ and $\kappa$ and, therefore, the effective optical rotation $\theta_{\rm{eff}}$ and ellipticity $\eta_{\rm{eff}}$ per unit length are given by: $\theta_{\rm{eff}}(\omega) = (\omega/c) \rm{Re}(\kappa+\kappa_s)$, $\eta_{\rm{eff}}(\omega) = (\omega/c) \rm{Im}(\kappa+\kappa_s)$ and are shown in Fig.\,\,\ref{fig:absolute}.\\
\begin{figure}[t!]
		\includegraphics[width=\linewidth]{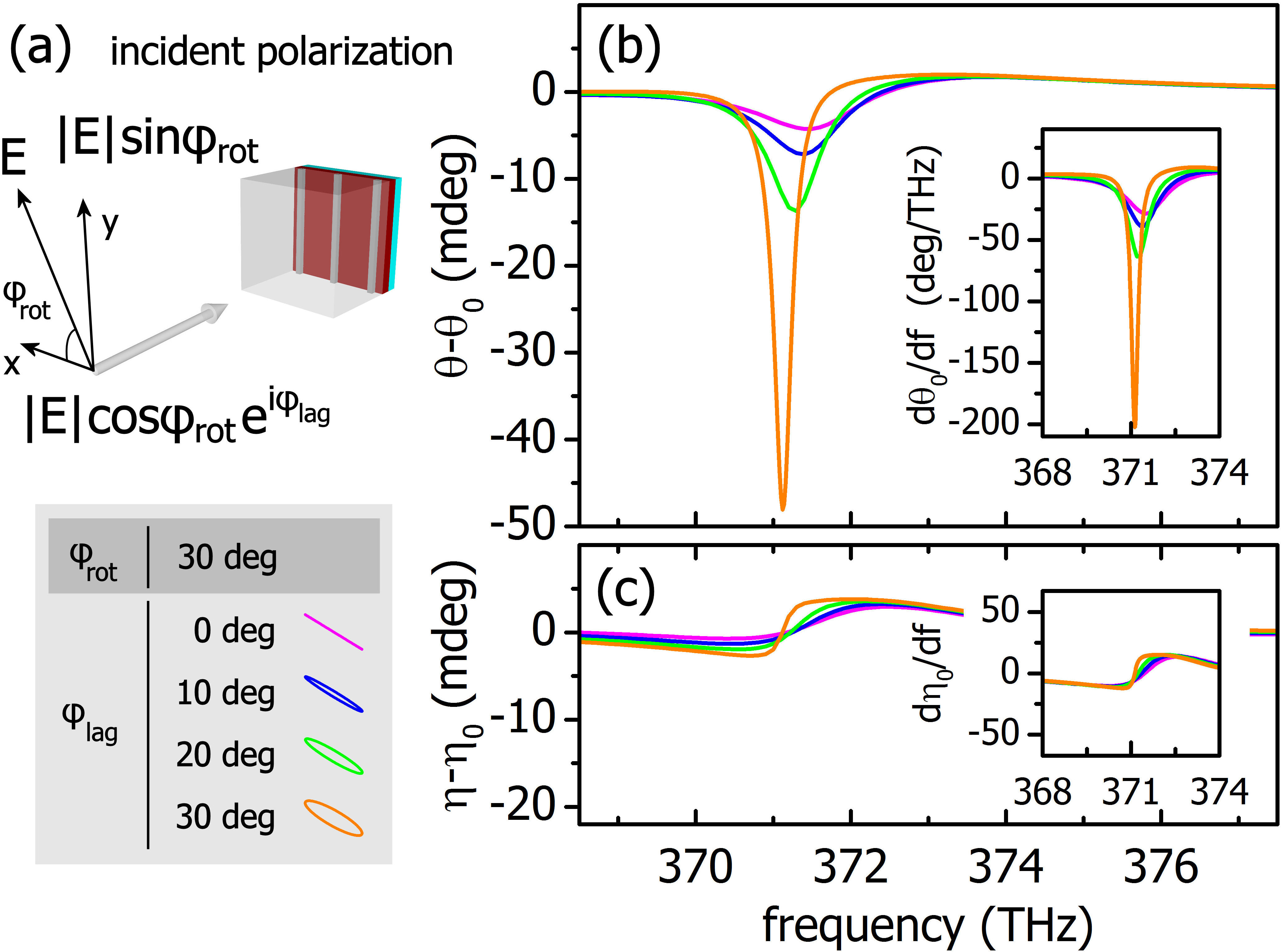}\vspace*{-8pt}
	\caption{\small{Enhanced chiral sensing using elliptically polarized incident wave. (a) Incident polarization and polarization ellipse for selected examples. Subtraction of far-field measurements of optical rotation and ellipticity in the presence and absence of the chiral layer, i.e. (b) $\theta-\theta_0$, and (c) $\eta-\eta_0$, for the selected polarizations shown in (a). For $\phi_{\rm{rot}} = 30$\,deg,  $\phi_{\rm{lag}} = 30$\,deg, the optical rotation signal is enhanced by a factor of $\sim200$ compared to the respective signal of the pure chiral film.}}
    	\label{fig:absolutePhirot}
\end{figure}
\indent Using our metasurface for chiral sensing and aligning the incident polarization to be either parallel or vertical to the metal wires, we obtain enhanced chiroptical signals and avoid any background contribution from the photonic structure. Indeed, for $\kappa$=0 (in the simulations and the analytical model) we obtain $\theta$=0 and $\eta$=0. However, in these cases the maximum enhancement is bound by the modal Q factors. To further enhance the far-field signals we can use an elliptically polarized incident wave to excite the metasurface's modes coherently and exploit the maximum attainable Q factor. However, subtraction of background signals would be required because the system is strongly anisotropic and the measured signals unavoidably contain the metasurface's response. Experimentally, one approach to remove the background contribution is to perform measurements with and without the chiral layer (similarly to the work of Ref.\,\cite{Zhao2017}), and subtracting the signals to yield the chiroptical signal. In Fig.\,\ref{fig:absolutePhirot} we show this possibility, where we choose an elliptically polarized incident wave $E = (E_x, E_y)\!=\!(E_0 \cos(\phi_{\rm{rot}}) e^{i\phi_{\rm{lag}}}, E_0 \sin(\phi_{\rm{rot}}))$ to excite the metamaterial. The angle $\phi_{\rm{rot}}$ is the angle between the incident wave's $E$-field and the x-axis, and $\phi_{\rm{lag}}$ tunes the phase-lag between the $x$,\,$y$ wave components; $E_0$ is a complex constant. For fixed $\phi_{\rm{rot}}$=\,30\,deg and increasing $\phi_{\rm{lag}}$ we find that the difference between two measurements with and without the chiral layer, i.e. $\theta-\theta_0$ [where $\theta$ ($\theta_0$) is with (without) the chiral layer], increases and becomes maximum for $\phi_{\rm{lag}}$\,=\,30\,deg, for which we obtain enhanced chiroptical signals by a factor of $\sim$200 compared to the respective signal from the pure chiral film ($\sim$0.24\,mdeg, see Fig.\,2). Meanwhile, the corresponding measurement for $\eta$ provides enhancement by a factor of $\sim$30, which naturally raises the question of whether the enhancement mechanism is related to some process other than the mode coupling. To answer this question, we expand $\theta\equiv\theta(f,\kappa)$ (a function of the frequency $f$ and the chirality parameter $\kappa$), in terms of $\kappa$. If we keep the first two terms we find $\theta-\theta_0\approx\partial\theta/\partial\kappa\rvert_{\kappa=0}\cdot\kappa$ or $\theta-\theta_0\approx(\partial\theta/\partial f)(\partial f/\partial\kappa)\rvert_{\kappa=0}\cdot\kappa$ where $\theta_0=\theta(f,0)$. Because $\theta, \eta$ scale linearly with $\kappa$ (see SI), we can write $\partial \theta/\partial f\propto d\theta_0/df$, which leads to $\theta-\theta_0 \sim (d\theta_0/df)(df/d\kappa)\cdot\kappa$ (similarly we obtain $\eta-\eta_0 \sim (d\eta_0/df)(df/d\kappa)\cdot\kappa$). This result implies that for a certain $\kappa$, the far-field differential signals ($\theta-\theta_0$, $\eta-\eta_0$) are expected to be enhanced at frequencies where the background birefringence of the metasurface ($\theta_0$, $\eta_0$) is strongly dispersive. To independently verify this we remove the chiral layer and measure the signals $\theta_0$ and $\eta_0$. In Fig.\,\ref{fig:absolutePhirot} (insets) we present the calculated derivatives $d\theta_0/df$ and $d\eta_0/df$, which clearly follow $\theta-\theta_0$ and $\eta-\eta_0$, respectively. Thus, the chiroptical enhancement is mediated by the strongly dispersive metasurface's birefringence. This is in contrast to most contemporary approaches, which are based on the excitation of superchiral near-fields and the background birefringence is undesired\,\cite{Mohammadi2018,Solomon2019,Mohammadi2019}. Hence, turning what is typically viewed as a negative aspect to an important advantage, we achieve an enhancement of 2 orders of magnitude. In our case, the generation of superchiral near-fields, although beneficial, is not a prerequisite (see SI).\\
\begin{figure}[t!]
		\includegraphics[width=\linewidth]{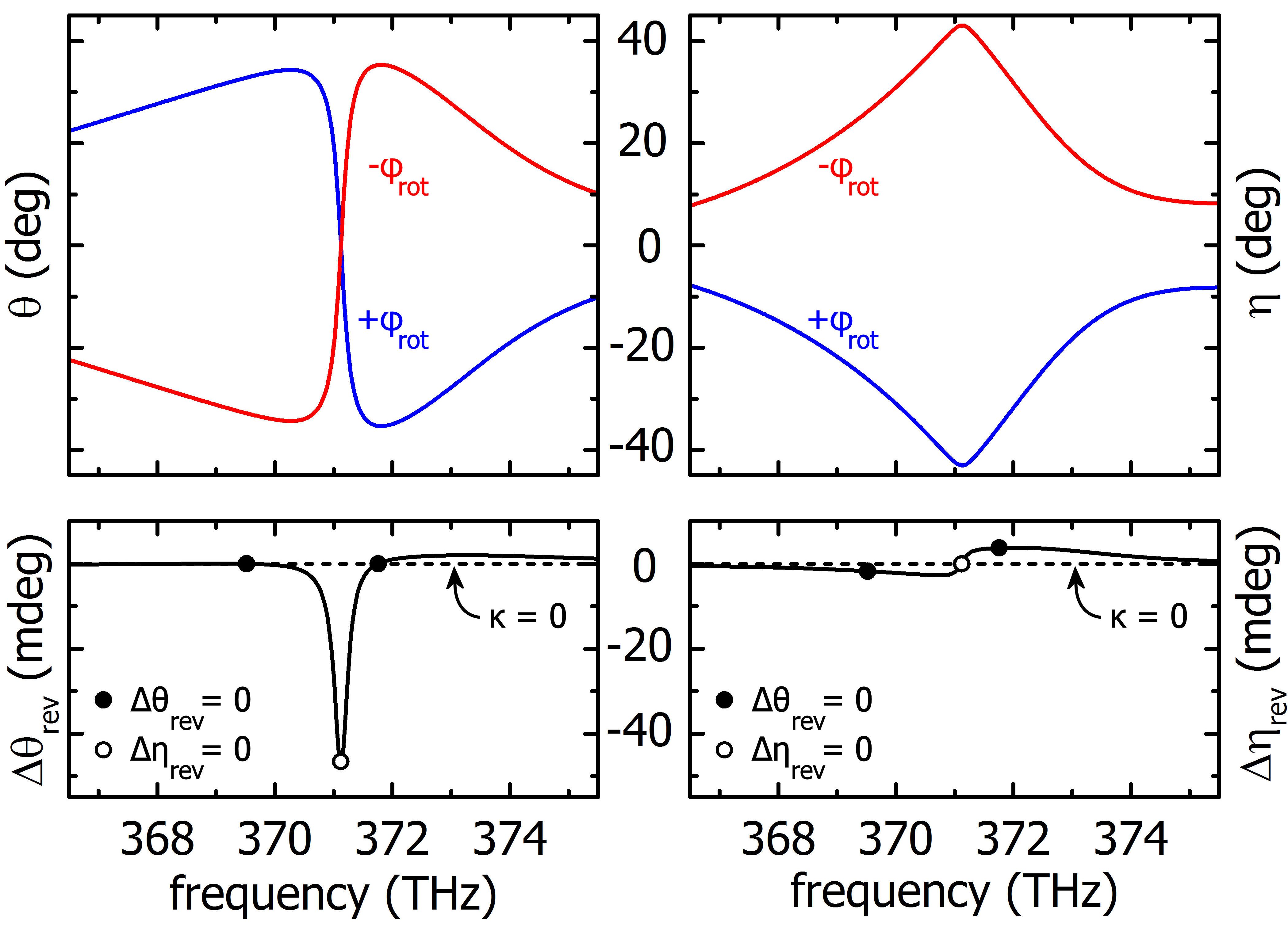}\vspace*{-8pt}
	\caption{\small{Absolute measurement of $\kappa$ with polarization reversal of elliptically polarized incident wave. Measurements for $\phi_{\rm{lag}} =30$\,deg with $\phi_{\rm{lag}} =\pm30$\,deg (blue and red lines, respectively, top row) are averaged to yield pure chiral signals $\Delta\theta_{\rm{rev}}$, $\Delta\eta_{\rm{rev}}$ (black lines, bottom row). Solid lines: $\kappa=+10^{-5}$. Dashed lines: $\kappa = 0$. Frequencies of pure optical rotation and pure ellipticity are marked with open dots and filled dots, respectively.}}
    	\label{fig:reversal}
\end{figure}
\indent Notwithstanding, our system offers the possibility for a crucial signal reversal with which one can directly isolate the enhanced chiroptical signal without the need for sample removal and interference with the system, a unique approach in metamaterial-based chiral sensing schemes: excitation with reversed polarization yields separate polarization effects of opposite sign, thus enabling the isolation of signals originating only from the chiral medium. To illustrate this, in Fig.\,\ref{fig:reversal} we present simulations of $\theta$ and $\eta$ for incident elliptical waves with $\phi_{\rm{rot}}= \pm30$\,deg and $\phi_{\rm{lag}} = 30$\,deg, which acquire (opposite) rotations and ellipticities that we label as $\theta_\pm$ and $\eta_\pm$. By taking their average, i.e. $2\Delta\theta_{\rm{rev}} \equiv \theta_++\theta_-$ and $2\Delta\eta_{\rm{rev}} \equiv \eta_++\eta_-$, any signal originating from the metasurface is cancelled (and similarly other potential achiral backgrounds), while the pure chiroptical signal, which is even under this polarization reversal, doubles. Our results are also in agreement with the measurements we present in Fig.\,\ref{fig:absolutePhirot}. Most importantly, we observe that for specific frequencies we obtain pure, enhanced, rotation signals ($\eta\simeq0$, e.g. at $\sim$371\,THz), while for other frequencies $\theta\simeq0$, $\eta\neq0$ (see Fig.\,\ref{fig:absolute}), enabling, thus, practical enhanced absolute polarimetric measurements of the chiral refractive index. The importance of the signal reversal becomes apparent: under realistic experimental conditions, one can appropriately tune the frequency of the probing radiation around the resonance of the metamaterial and apply this reversal (e.g. with the use of polarization modulators), isolating the chiral signal even under the presence of high-noise environments and other achiral effects, similarly to how signal reversals have allowed for chiral sensing in conditions where traditional polarimetry fails to perform\,\cite{Sofikitis2014,Bougas2015}.\\
\indent In this Letter we demonstrated a dielectric metamaterial system that provides a platform where aspects crucial for chiral sensing in the nanoscale can be realized: (a) absolute measurements of chirality, (b) signal enhancements of at least 2 orders of magnitude stronger than typical polarimetric techniques, and (c) signal reversals for background signal cancellation. With signal reversals, in particular, the measurement can be performed with improved signal-to-noise ratio, as the signals are twice as strong, and modulation techniques can be employed\,\cite{Sofikitis2014,Bougas2015}. Additionally, compared to chiral sensing approaches that employ nanostructures of complicated geometry, our system offers a large non-structured surface on which the chiral sample can be placed, without the need for special pursuit of hot spots\,\cite{Hendry2010,Tullius2015,Tullius2017,Zhao2017} or the requirement for embedding the optically active material inside the photonic structure\,\cite{Mohammadi2019}.\\
\indent This work was supported by the European Commission Horizon 2020, project ULTRACHIRAL (Grant No. FETOPEN-737071). SD thanks T. Koschny, and LB thanks G. Iwata and A. Harding for fruitful discussions. 

\bibliography{AbsCHIbib}

\end{document}